\documentclass[preprint,endfloats*,superscriptaddress,showkeys,showpacs]{revtex4-1}

\usepackage{graphicx}
\usepackage{dcolumn}
\usepackage{bm}
\usepackage{amsmath}
\usepackage{latexsym}
\usepackage{hyperref} 
\usepackage{color,soul} 
\usepackage{CJKutf8}

\newcommand{\rfig}[1]{Fig.~\ref{#1}}

\newcommand{\req}[1]{Eq.~(\ref{#1})}

\makeindex

\begin{document}

\title{On the Onsager{\textendash}Casimir reciprocal relations in a tilted Weyl semimetal}

\author{Bingyan Jiang(\begin{CJK}{UTF8}{gbsn}江丙炎\end{CJK})}
\author{Jiaji Zhao(\begin{CJK}{UTF8}{gbsn}赵嘉佶\end{CJK})}
\author{Lujunyu Wang(\begin{CJK}{UTF8}{gbsn}王陆君瑜\end{CJK})}
\author{Ran Bi(\begin{CJK}{UTF8}{gbsn}毕然\end{CJK})}
\author{Juewen Fan(\begin{CJK}{UTF8}{gbsn}范珏雯\end{CJK})}
\affiliation{State Key Laboratory for Artificial Microstructure and Mesoscopic Physics, Frontiers Science Center for Nano-optoelectronics, Peking University, Beijing 100871, China}
\author{Zhilin Li(\begin{CJK}{UTF8}{gbsn}李治林\end{CJK})}
\affiliation{Beijing National Laboratory for Condensed Matter Physics, Institute of Physics, Chinese Academy of Sciences, Beijing 100190, China}
\author{Xiaosong Wu(\begin{CJK}{UTF8}{gbsn}吴孝松\end{CJK})}
\email{xswu@pku.edu.cn}
\affiliation{State Key Laboratory for Artificial Microstructure and Mesoscopic Physics, Frontiers Science Center for Nano-optoelectronics, Peking University, Beijing 100871, China}
\affiliation{Collaborative Innovation Center of Quantum Matter, Beijing 100871, China}

\begin{abstract}
The Onsager{\textendash}Casimir reciprocal relations are a fundamental symmetry of nonequilibrium statistical systems. Here we study an unusual chirality-dependent Hall effect in a tilted Weyl semimetal Co$_3$Sn$_2$S$_2$ with broken time reversal symmetry. It is confirmed that the reciprocal relations are satisfied. Since two Berry curvature effects, an anomalous velocity and a chiral chemical potential, contribute to the observed Hall effect, the reciprocal relations suggest their intriguing connection.
\end{abstract}



\maketitle

Onsager's reciprocity relations state that in thermodynamic systems out of equilibrium, the linear response coefficient, linking a thermodynamic driving force and the resultant flow, is a symmetric tensor. That is, ${\cal L}_{ij}={\cal L}_{ji}$. In the presence of time reversal breaking fields, it becomes the Onsager{\textendash}Casimir reciprocal relations, ${\cal L}_{ij}({\cal B})={\cal L}_{ji}(-{\cal B})$, where ${\cal B}$ denotes all time reversal breaking fields. These relations, rooted in the principle of microscopic reversibility, are a fundamental symmetry of nonequilibrium statistical systems. When multiple coupled pairs of flows and thermodynamic forces are involved, the relations can offer insights into the connection between different response coefficients that seem, at first glance, unrelated. For instance, in his seminal papers in 1931, Onsager derived the relations for several coupled irreversible processes and related different coefficients to one another\cite{Onsager1931,Onsager1931Dec}. The reciprocal relations have also played an important role on understanding various spin-related transport\cite{Jacquod2012}. 

Recently, an unusual chirality-dependent Hall effect and magnetoresistance that are antisymmetric in both magnetic field and magnetization were predicted to occur in tilted topological Weyl semimetals\cite{Zyuzin2017,Sharma2017,Wei2018,Ma2019a,Das2019,Johansson2019,Kundu2020}. It was suggested that such an antisymmetry violates the Onsager{\textendash}Casimir reciprocal relations\cite{Das2019}. Soon, the antisymmetric Hall and magnetoresistance were experimentally observed in a magnetic Weyl semimetal, Co$_3$Sn$_2$S$_2$\cite{Jiang2021Jun}. The phenomenon is originated from the Berry curvature of energy bands and involves several effects. In particular, the longitudinal linear magnetoresistance results from the Berry curvature correction to the phase space volume, while the Hall effect consists of contributions from the chiral chemical potential and the anomalous velocity. Here, we study the chirality-dependent Hall effect in Co$_3$Sn$_2$S$_2$, with a focus on the validity of the Onsager{\textendash}Casimir reciprocal relations. Two methods were employed and it was found that the reciprocal relations hold for this Hall effect. The relations connect the effect of the chiral chemical potential to that of the anomalous velocity. Despite only one pair of flow and force at work, our results reveal an intriguing connection between two effects, which deepens our understanding of Berry-curvature-related electrical transport.

\begin{figure}[htbp]
	\begin{center}
		\includegraphics[width=1\columnwidth]{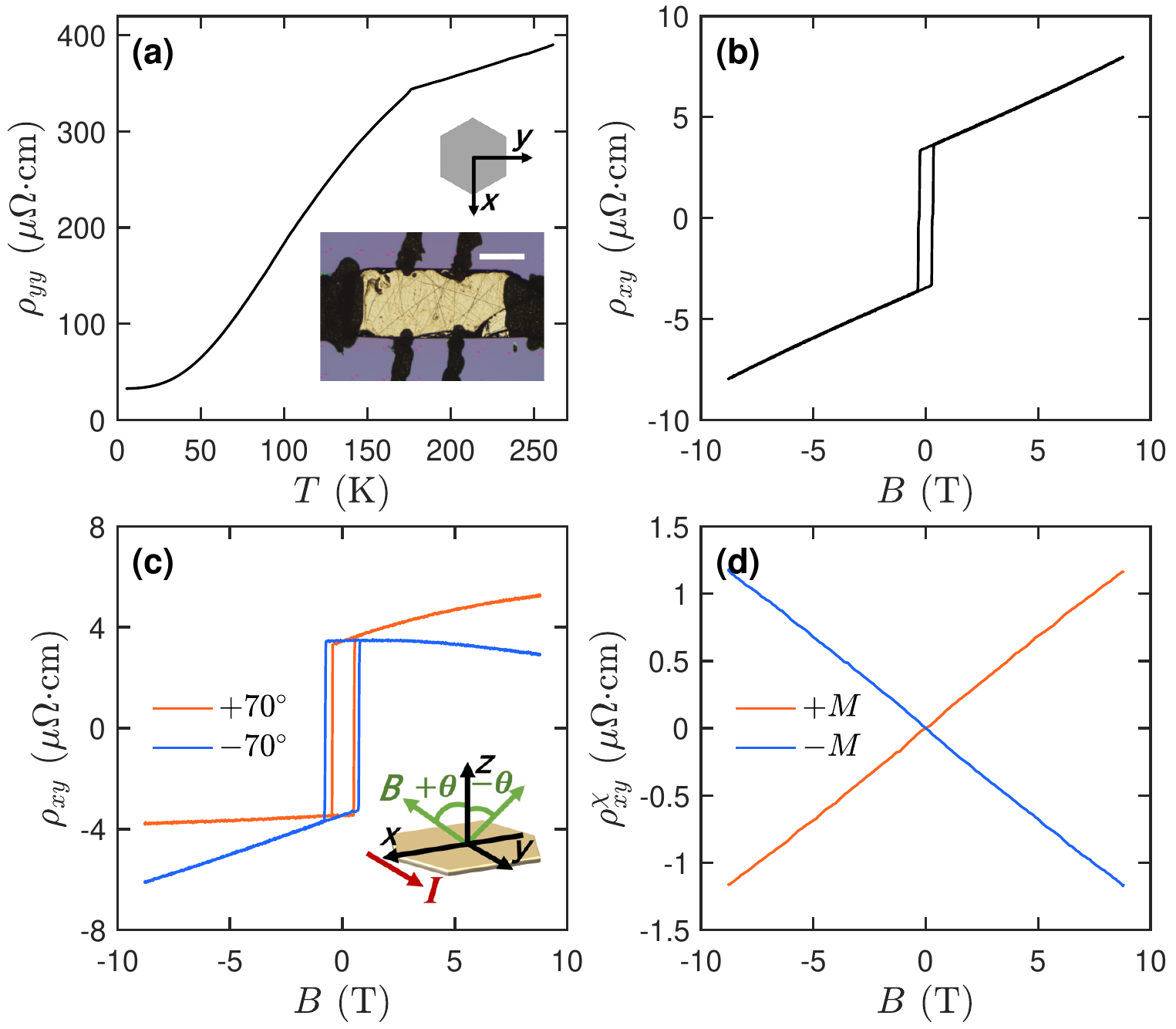}
		\caption{Basic transport properties of Co$_3$Sn$_2$S$_2$ Sample \#1 and the chirality-dependent Hall effect. \textbf{a}, Temperature dependence of the longitudinal resistivity $\rho_{yy}$. A clear kink indicates a Curie temperature of 175 K. The inset shows the Hall bar device and the scale bar is 300 $\mu$m. \textbf{b}, Anomalous Hall effect at $T= 50$ K. The linear dependence of $\rho_{xy}$ on the magnetic field above $B_\mathrm{c}$ suggests a single carrier type. \textbf{c}, Hall resistivity as the magnetic field scans along $\pm70^\circ$ in the $xz$ plane at $T= 50$ K. The inset depicts the measurement setup and the magnetic field direction. \textbf{d}, The chirality-dependent Hall resistivity $\rho_{xy}^\chi$ obtained by symmetrizing the data in \textbf{c}. $\rho_{xy}^\chi$ is linear in magnetic field and changes its sign as the magnetization reverses.}
		\label{FigBasic}
	\end{center}
\end{figure}

Co$_3$Sn$_2$S$_2$ single crystals were grown by the chemical vapor transport method\cite{Jiang2021Jun} and polished to a bar shape for the transport measurements. We utilized silver paste to fabricate the electrical contacts. Before applying the paste, the crystal was treated with Ar plasma to improve the contact.  The device is shown in the inset of \rfig{FigBasic}a and the $x$ axis in the diagram is along the crystal [100] direction. The electrical transport measurements were performed using a low frequency lock-in method. A typical current of 2 {\textendash} 3 mA was employed. The angular dependence experiment was carried out using a motorized rotator stage with a resolution of 0.02$^\circ$. Our samples exhibit transport properties similar to previous reports\cite{Yang2020c,Jiang2021Jun,Wang2018,Liu2018}. The temperature dependence of the longitudinal resistivity $\rho_{yy}$ is shown in \rfig{FigBasic}a, where a ferromagnetic phase transition manifests as a clear kink at the Curie temperature $T_\mathrm{c} \sim 175$ K. Below $T_\mathrm{c}$, the anomalous Hall effect is observed, as shown in \rfig{FigBasic}b. The Hall resistivity $\rho_{xy}$ jumps at the coercive field $B_\mathrm{c}$ and shows a linear dependence on the magnetic field above $B_\mathrm{c}$, from which we obtain an ordinary Hall coefficient $R_\mathrm{H}=0.51\ \mu\Omega\cdot$cm/T and a corresponding carrier density of $1.2\times10^{21}$ cm$^{-3}$.

Co$_3$Sn$_2$S$_2$ is a magnetic Weyl semimetal\cite{Liu2018,Liu2019,Morali2019}. The interplay of band topology and magnetism gives rise to a number of unusual phenomena, attracting intensive research interest\cite{Wang2018,Liu2018,Liu2019,Morali2019,Guin2019a,Shen2019,Yin2019,Geishendorf2019Mar,Yang2020b,Tanaka2020Oct,Shama2020,Lachman2020,Yang2020c,Shen2020a,Xu2020a,Geishendorf2020Jan,Ding2021Aug,Howard2021Jul,Zhang2021Sep}. A very recent experiment has uncovered a chirality-dependent Hall effect that is antisymmetric in both the in-plane magnetic field and the magnetization\cite{Jiang2021Jun}. It is believed to be associated with the titled Weyl cones. The chirality-dependent Hall effect is observed in our samples. When the magnetic field is rotated from the $z$ axis to the $x$ axis, as depicted in the inset of \rfig{FigBasic}c, the in-plane field generates an extra Hall resistivity proportional to $B_x$. Therefore, the transverse signals at $\pm\theta$ diverge, shown in \rfig{FigBasic}c ($\theta=\pm70^\circ$ for instance). The chirality-dependent Hall effect $\rho_{xy}^\chi$ can be obtained by data antisymmetrizing as $\rho_{xy}^\chi(\theta)=[\rho'_{xy}(+\theta)-\rho'_{xy}(-\theta)]/2$, where $\rho'_{xy}=\rho_{xy}-\rho_{xy}^\mathrm{AHE}$. Considering that the amplitude and orientation of the magnetization remain almost unchanged at 50 K due to a strong magnetocrystalline anisotropy\cite{Shen2019,Guin2019a}, a constant anomalous Hall resistivity $\rho_{xy}^\mathrm{AHE}$ is subtracted. The obtained $\rho_{xy}^\chi$ for positive and negative magnetization at $\theta=\pm70^\circ$ are plotted in \rfig{FigBasic}d, which is consistent with the previously reported results of $\rho_{xy}^\chi\propto B_xM_z$\cite{Jiang2021Jun}.

\begin{figure}[htbp]
	\begin{center}
		\includegraphics[width=1\columnwidth]{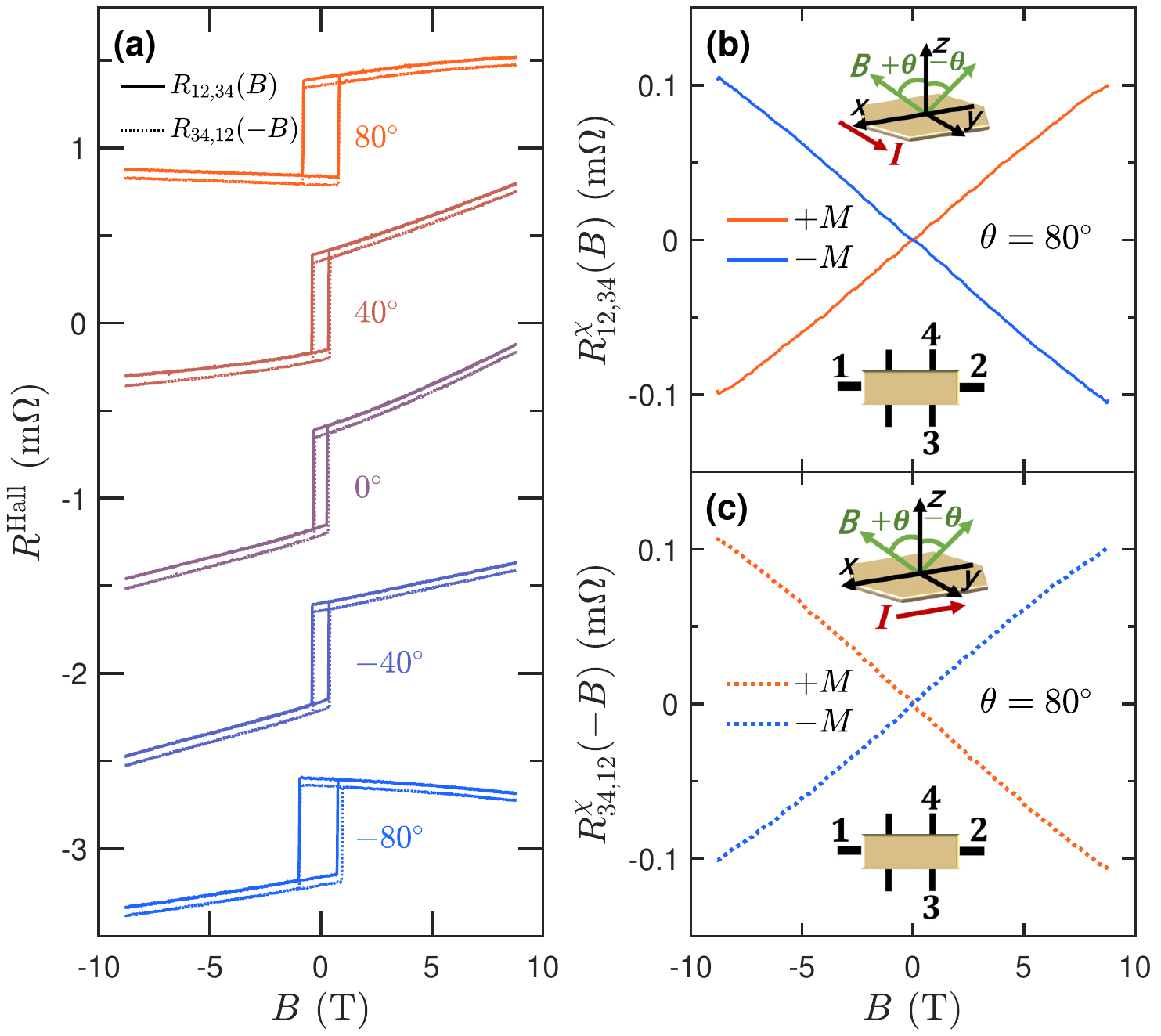}
		\caption{Verification of the reciprocal theorem, \req{eq.global.sym}. \textbf{a}, $R_{12,34}(B)$ and $R_{34,12}(-B)$ at different angles in the $xz$ plane for $T=$ 50 K. Data are shifted for clarity. \textbf{b} and \textbf{c}, $R_{12,34}^\chi(B)$ and $R_{34,12}^\chi(-B)$ at 80$^\circ$ by data antisymmetrizing, respectively. The bottom inset depicts the measurement configuration, while the top inset shows the angle definition. }
		\label{FigRB}
	\end{center}
\end{figure}

\begin{figure}[htbp]
	\begin{center}
		\includegraphics[width=1\columnwidth]{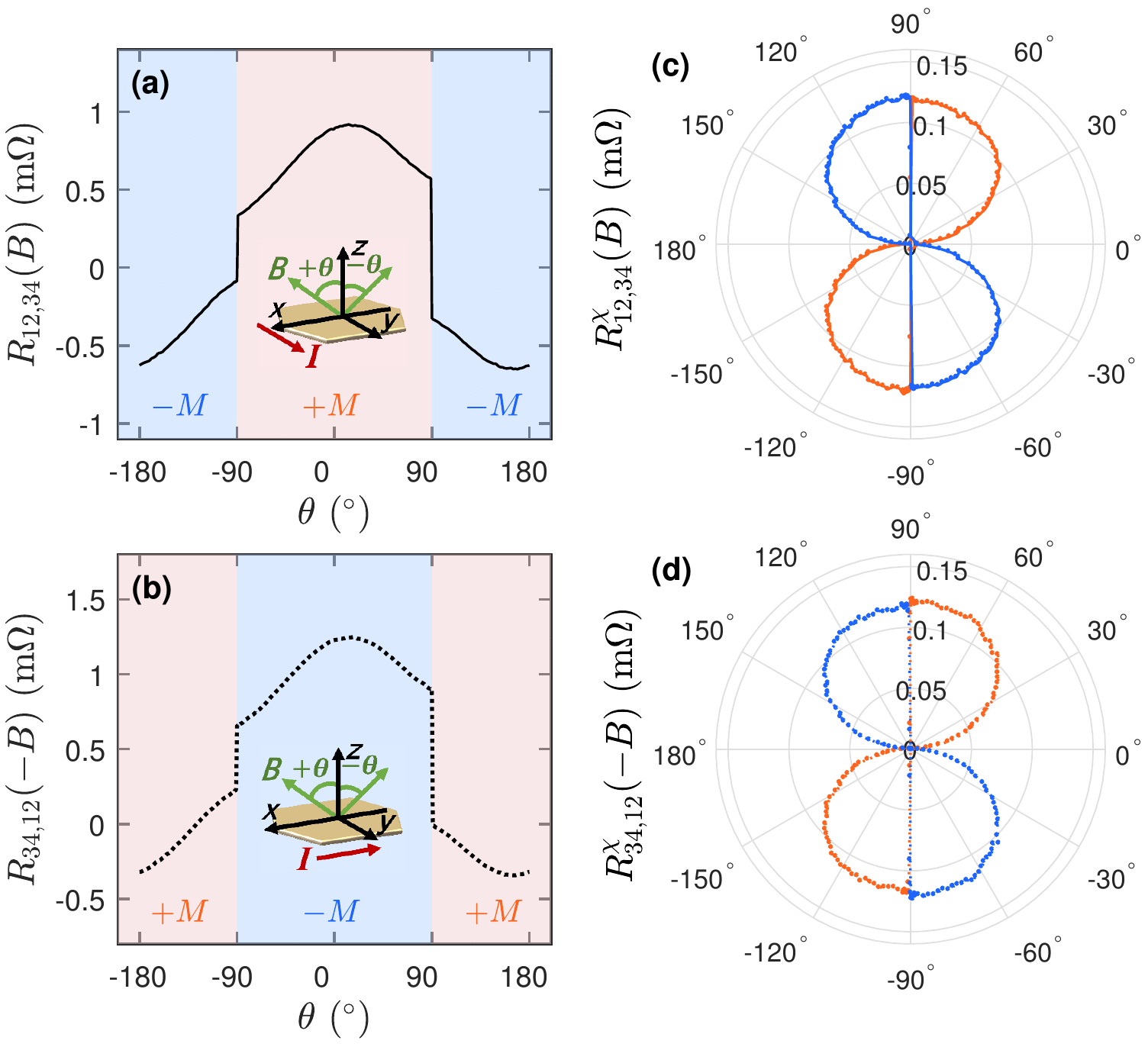}
		\caption{Angular dependence of the Hall resistance. \textbf{a} and \textbf{b}, Angular dependence of $R_{12,34}(B)$ and $R_{34,12}(-B)$ at $B= 8.8$ T for $T= 50$ K. \textbf{c} and \textbf{d}, Angular dependence of $R_{12,34}^\chi(B)$ and $R_{34,12}^\chi(-B)$ by data antisymmetrizing. The same angular dependence of $R$ and $R^\chi$ under two configurations further verifies the reciprocal theorem. }
		\label{FigRPhi}
	\end{center}
\end{figure}

The chirality-dependent Hall effect satisfies $\rho_{xy}(B,M)=\rho_{xy}(-B,-M)$, which is consistent with the Onsager{\textendash}Casimir reciprocal relations\cite{Jiang2021Jun}. However, to prove that the reciprocal relations hold, one needs to verify $\rho_{xy}(B,M)=\rho_{yx}(-B,-M)$. One method is to attest the reciprocity theorem
\begin{equation}
R_{12,34}(B,M)=R_{34,12}(-B,-M),
\label{eq.global.sym}
\end{equation}
where the two pairs of indices denote the current leads and voltage leads, respectively\cite{Buttiker1986Oct,Buttiker1988}. \req{eq.global.sym} can be derived from the Onsager{\textendash}Casimir reciprocal relations. \rfig{FigRB} shows the Hall effect measured under two configurations. In the first configuration, the current is applied through contacts 1 and 2, and the voltage between 3 and 4 is measured to obtain $R_{12,34}$, as depicted in the bottom inset of \rfig{FigRB}b. In the second configuration, the current and voltage leads are swapped. The current is applied through contacts 3 and 4, and the voltage between 1 and 2 is measured to obtain $R_{34,12}$, as depicted in the bottom inset of \rfig{FigRB}c. Measurements were performed at several angles. Data are shown in \rfig{FigRB}a. In all cases, $R_{34,12}(-B,-M)$ follows $R_{12,34}(B,M)$ exactly. Consequently, the extracted chirality-dependent Hall effect satisfies the reciprocity theorem, $R^\chi_{12,34}(B,M)=R^\chi_{34,12}(-B,-M)$, as shown in \rfig{FigRB}b and c. We further measured the magnetic field angular dependence of two resistances, shown in \rfig{FigRPhi}a and b. The magnetic field was rotated in the $xz$ plane. The chirality-dependent Hall resistivity can be extracted as $R^\chi(\theta)=[R(+\theta)-R(-\theta)]/2$ and is presented in \rfig{FigRPhi}c and d. The angular dependences of $R_{12,34}^\chi(B,M)$ and $R_{34,12}^\chi(-B,-M)$ show almost the same behavior, further supporting the validity of \req{eq.global.sym}. 

However, \req{eq.global.sym} is a global symmetry. In principle, a whole series of four-probe measurements must be carried out and all results must comply with \req{eq.global.sym} in order to prove the validity of the Onsager{\textendash}Casimir relations\cite{Casimir1945Apr,Buttiker1988}. We now attempt another method, i.e., direct verification of the local symmetry of $\rho_{xy}(B,M)=\rho_{yx}(-B,-M)$. Note that there is a practical disadvantage in this method, as it is difficult to prepare Hall bars along two directions on the same crystal. In \rfig{FigRB}, $R_{12,34}(B,M)$, corresponding to $\rho_{xy}(B,M)$, was measured. We had to prepare a Hall bar with another sample to measure $\rho_{yx}(-B,-M)$. Since no two samples can have exactly the same resistivity, only a semi-quantitative comparison will be made.

\begin{figure}[htbp]
	\begin{center}
		\includegraphics[width=1\columnwidth]{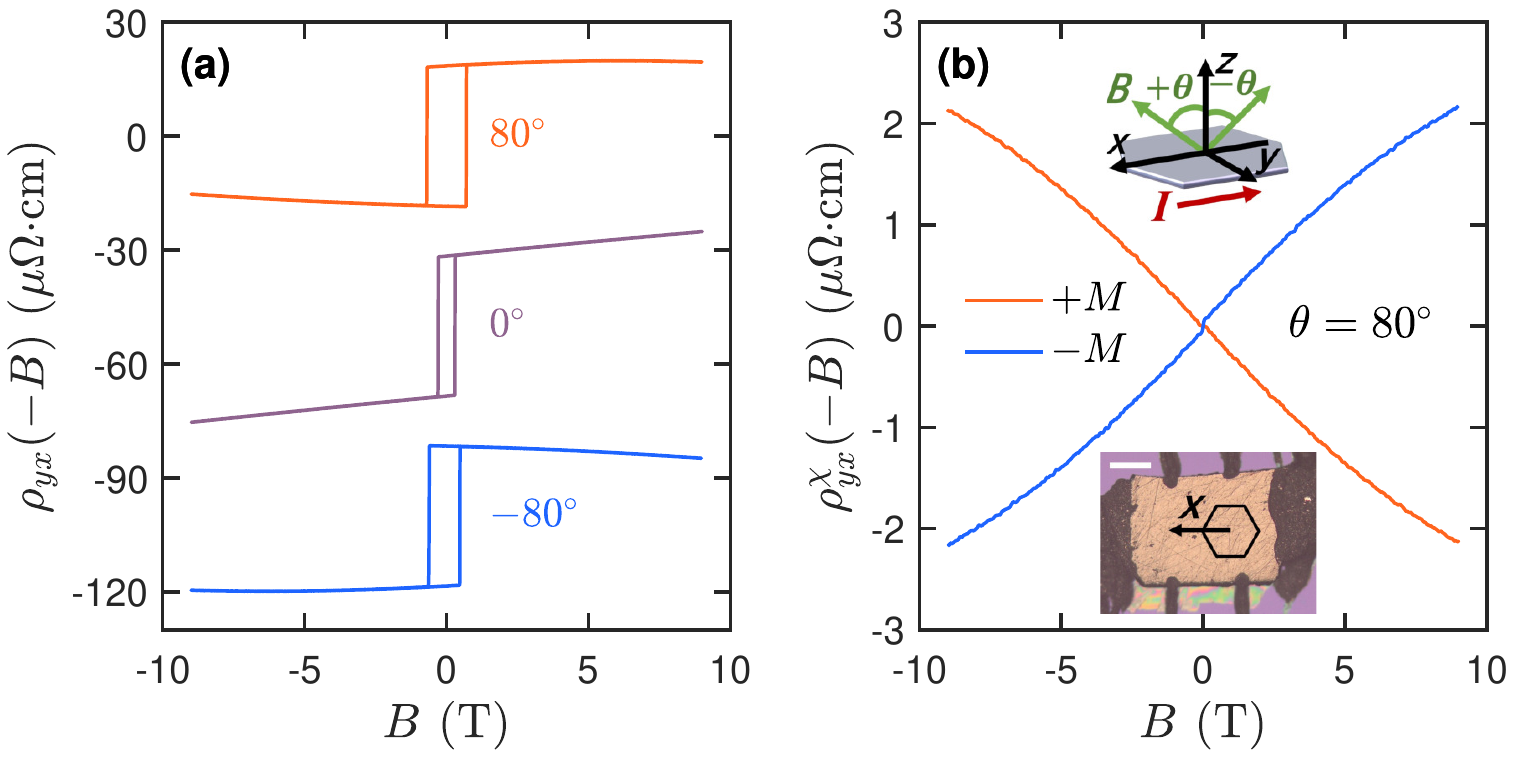}
		\caption{Verification of the local Onsager{\textendash}Casimir relations. \textbf{a}, $\rho_{yx}(-B)$ of Sample \#2 at several angles for $T= 50$ K. The behavior is similar to Sample \#1. Data are shifted for clarity. \textbf{b}, $\rho_{yx}^\chi(-B)$ at 80$^\circ$ for positive and negative magnetization. The top inset depicts the measurement setup. The bottom inset shows an optical image of Sample \#2 and the scale bar is 300 $\mu$m. }
		\label{FigB01}
	\end{center}
\end{figure}

\rfig{FigB01}a displays the measured $\rho_{yx}(-B)$ at $\theta=80^\circ, 0^\circ, -80^\circ$ for Sample \#2. Above $B_\mathrm{c}$, $\rho_{yx}$ exhibits a linear $B$-dependence, from which a carrier density of $8.4\times10^{20}$ cm$^{-3}$ is obtained. It is slightly smaller than the carrier density of Sample \#1. The resistivity at zero field is about 125 $\mu\Omega\cdot$cm, compared to 65 $\mu\Omega\cdot$cm of Sample \#1. With increasing magnetic field or angle $\theta$, $\rho_{yx}(-B)$ at $\pm \theta$ diverge. This behavior qualitatively agrees with that observed in Sample \#1. The chirality-dependent Hall resistivity $\rho^\chi_{yx}(-B)$ can be extracted by symmetrizing, depicted in \rfig{FigB01}b. It is antisymmetric in both magnetic field and magnetization, mimicking $\rho_{xy}(B,M)$, which confirms the local Onsager{\textendash}Casimir reciprocal relations. The amplitude of $\rho^\chi_{yx}$ for Sample \#2 is nearly twice as much as that for Sample \#1, which we believe can be attributed to the difference of a factor of 2 in the resistivity.

To reveal the implication of the reciprocal relations on the chirality-dependent Hall effect, it is necessary to identify various effects of the Berry curvature involved in the phenomenon. Start from the equations of motion for band electrons. The velocity can be expressed as
\begin{equation}
\bm{\dot{r}}=D(\bm{B},\bm{\Omega_k})\left[ \bm{v_k}+\frac{e}{\hbar}\bm{E} \times \bm{\Omega_k}+\frac{e}{\hbar}(\bm{v_k\cdot \Omega_k})\bm{B} \right]
\label{eq.velocity}
\end{equation}
, where $\bm{v_k}=\frac{1}{\hbar} \frac{\partial\epsilon}{\partial\bm{k}}$ is the band group velocity, $\bm{\Omega_k} \propto \chi \frac{\bm{k}}{k^3}$ is the Berry curvature, and $D(\bm{B},\bm{\Omega_k})=[1+\frac{e}{\hbar}(\bm{B\cdot\Omega_k})]^{-1}$ is the modification of the phase space volume\cite{Xiao2010,Xiao2005}. $e$, $\hbar$, $\bm{E}$ and $\bm{B}$ are the elementary charge, the reduced Planck constant, electric field and magnetic field, respectively. $\chi=\pm 1$ is the chirality of Weyl cones. $\frac{e}{\hbar}(\bm{v_k\cdot \Omega_k})\bm{B}$ is the anomalous velocity related to the Berry curvature. One can calculate the nonequilibrium distribution function by solving the Boltzmann equation under a relaxation time approximation and arrive at\cite{Ma2019a,Das2019} 
\begin{equation}
f_{\bm{k}}=f_0+\left[ eD\tau\bm{E\cdot v_k}+ \frac{e^2}{\hbar}D\tau(\bm{B\cdot E})(\bm{v_k\cdot \Omega_k})\right] \times \left( -\frac{\partial f_0}{\partial \epsilon_{\bm{k}}} \right)
\label{eq.distr}
\end{equation}
, neglecting higher-order corrections. Here, $f_0$ is the equilibrium distribution function, and $\tau$ is the mean free time. The first term in the square bracket is the shifted distribution induced by the electric field, while the second term is the chiral chemical potential, resulting from pumping of electrons between Weyl cones of opposite chiralities. The current can be calculated by integration of the velocity
\begin{equation}
\bm{j}=-e\int \frac{\mathrm{d}\bm{k}}{(2\pi)^3} D^{-1}\bm{\dot{r}}f_{\bm{k}}.
\label{eq.current}
\end{equation}
When calculating the integral, it should be born in mind that Weyl cones appear in pairs. There are three contributions that are linear in magnetic field and electric field, which are non-zero only when Weyl cones are tilted. One is the product of the first term of \req{eq.velocity} and the second term in \req{eq.distr}. It generally contributes to the diagonal current and hence is not of our interest. We write down the other two contributions as
\begin{equation}
\bm{j}=-\frac{e^3\tau}{\hbar}\int \frac{\mathrm{d}\bm{k}}{(2\pi)^3} D \left[ (\bm{E\cdot v_k})(\bm{v_k\cdot \Omega_k})\bm{B} + (\bm{B\cdot E})(\bm{v_k\cdot \Omega_k})\bm{v_k} \right] \times \left( -\frac{\partial f_0}{\partial \epsilon_{\bm{k}} } \right).
\label{eq.off-diagonal}
\end{equation}
The first current is the contribution of the anomalous velocity from the shifted distribution, denoted as $\bm{j}_\mathrm{A}$. The second current is the contribution of the velocity from the chiral chemical potential, denoted as $\bm{j}_\mathrm{C}$.

\begin{figure}[htbp]
	\begin{center}
		\includegraphics[width=1\columnwidth]{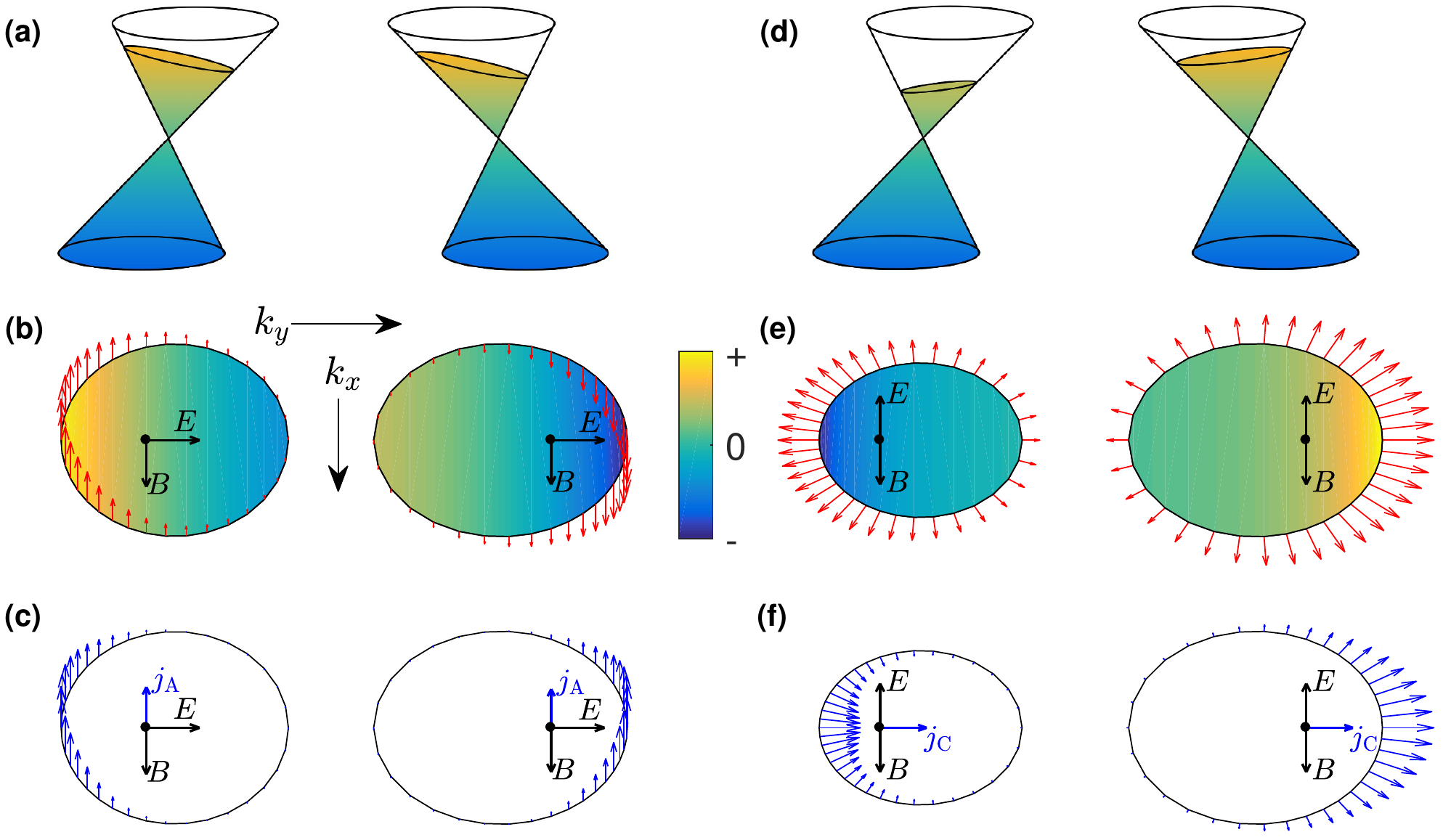}
		\caption{Schematic diagram of different contributions to the $B$-linear transverse current. \textbf{a}, Weyl cones of $\chi=\pm1$ tilt along the $y$ direction. The Fermi level tilts due to the electric field. \textbf{b}, Anomalous velocity $\bm{v_a}\propto (\bm{v_k}\cdot\bm{\Omega_k})\cdot\bm{B}$ (red arrows) and the distribution function shift $\Delta f_\mathrm{E}\propto \bm{v_k}\cdot\bm{E}$ (contour plot) due to $\bm{E}$ field, where $\bm{E}\parallel y$ and $\bm{B}\parallel x$. \textbf{c}, $B$-linear current due to $\bm{v_a}\Delta f_\mathrm{E}$ (blue arrows on the Fermi surface), the integration of which over Fermi surfaces yields $\bm{j}_\mathrm{A}$. \textbf{d}, The influence of the chiral chemical potential on the Fermi level. The Fermi level also tilts as $\bm{v_k}\cdot\bm{\Omega_k}$ varies on the Fermi surface. \textbf{e}, $\bm{v_k}$ (red arrows) and the influence of the chiral chemical potential on the distribution function $\Delta f_\mathrm{ca}\propto (\bm{B}\cdot\bm{E})(\bm{v_k}\cdot\bm{\Omega_k})$ (contour plot) on the Fermi surface, where $\bm{E}\parallel -x$ and $\bm{B}\parallel x$. \textbf{f}, $B$-linear current due to $\bm{v_k}\Delta f_\mathrm{ca}$ (blue arrows).}
		\label{FigCal}
	\end{center}
\end{figure}

For simplicity, let us consider the particular configurations in our measurements. As depicted in \rfig{FigCal}, the tilt of Weyl cones is along the $y$ axis\cite{Jiang2021Jun}, while the in-plane magnetic field is along the $x$ axis. Various terms in the integrand are sketched to provide an intuitive estimation on the corresponding currents. For $\rho_{xy}$\cite{Note2}, the electric field is in the $y$ direction. $\bm{j}_\mathrm{C}$ vanishes because $\bm{B\cdot E}=0$. $\bm{j}_\mathrm{A}$ is parallel to $\bm{B}$ (transverse). Although the anomalous velocity $\frac{e}{\hbar}(\bm{v_k\cdot \Omega_k})\bm{B}$ are opposite in Weyl cones of opposite chiralities, the shifted distributions are not symmetric in the presence of a tilt, yielding a finite current. For $\rho_{yx}$, the electric field is in the $x$ direction. $\bm{j}_\mathrm{A}$ is in the $x$ direction and affects the longitudinal conductivity. The transverse component of current comes from $\bm{j}_\mathrm{C}$. It is worth pointing out that for Weyl cones without a tilt, when considering the effect of the chiral chemical potential, the Fermi level lies horizontally and only shifts in energy. In contrast, for tilted Weyl cones, the Fermi level not only shifts in energy, but also tilts along the $y$ direction, as sketched in \rfig{FigCal}d. When Weyl cones are not tilted, $\bm{j}_\mathrm{C}$ vanishes because $\bm{v_k}$ is odd in $\bm{k}$ within a Weyl cone. A tilt breaks the inversion symmetry of $\bm{v_k}$ within a Weyl cone and the balance between two cones, giving rise to a finite current. The current is along the direction of the tilt (the $y$ axis), as sketched in \rfig{FigCal}f. 

Our experiments indicate that the Onsager{\textendash}Casimir reciprocal relations hold for the chirality-dependent Hall effect, $\rho_{xy}(B,M) = \rho_{yx}(-B,-M)$. Based on the above analysis on the origin of $\rho_{xy}$ and $\rho_{yx}$, it can be therefore concluded that $\bm{j}_\mathrm{A}(B,M) = \bm{j}_\mathrm{C}(-B,-M)$ in our experiment configurations. Although these two currents are manifests of different effects, our results suggest that they are intricately connected.

Having verified the reciprocal relations under a particular electric and magnetic field configuration and pointed out its implication, we generalize the conclusion to all configurations. Substituting \req{eq.velocity} and \req{eq.distr} into \req{eq.current} and dropping the term for the anomalous Hall effect, one can express the conductivity tensor as
\begin{equation}
\sigma_{ij}=-e^2\tau \int \frac{\mathrm{d}\bm{k}}{(2\pi)^3} D \left[ v_i + \frac{e}{\hbar}(\bm{v_k\cdot \Omega_k})B_i \right] \left[ v_j + \frac{e}{\hbar}(\bm{v_k\cdot \Omega_k})B_j \right] \times \left( -\frac{\partial f_0}{\partial \epsilon_{\bm{k}} } \right),
\label{eq.generic}
\end{equation}
where $v_i$ denotes the $i$th component of $\bm{v_k}$. It was incorrectly concluded that \req{eq.generic} violates the reciprocal relations\cite{Das2019}. This misunderstanding stems from the implicit dependence of the anomalous velocity on the magnetization. In a time-reversal-symmetry-broken Weyl semimetal, reversing the magnetization leads to a reversal of the chiralities of Weyl nodes\cite{Wang2012}. Thus, the sign of $\bm{\Omega_k}$ depends on the sign of the magnetization. Taking the magnetization into consideration, \req{eq.generic} indeed satisfies the Onsager{\textendash}Casimir reciprocal relations, $\sigma_{ij}(B,M) = \sigma_{ji}(-B,-M)$. When $\bm{\Omega}=0$, \req{eq.generic} is reduced to the conventional formula. The effect of the Berry curvature is to introduce a correction (the anomalous velocity) to the velocity, besides the change of the phase space volume. 

\begin{acknowledgements}
We are grateful for discussions with J. Feng, J. R. Shi and H. Z. Lu. This work was mainly supported by National Key Basic Research Program of China (No. 2020YFA0308800) and NSFC (Project No. 11774009, No. 12074009). Z. L. Li acknowledges support from Beijing Natural Science Foundation (Project No. Z200008) and the Youth Innovation Promotion Association of CAS (No. 2021008).
\end{acknowledgements}

\providecommand{\newblock}{}


\end{document}